\documentclass[pra,aps,superscriptaddress,twocolumn]{revtex4}

\usepackage{graphicx}
\usepackage{amsmath}
\usepackage{bm}

\begin{document}

\title{Dynamical Casimir effect without boundary conditions}

\author{Hiroki Saito}
\affiliation{Interactive Research Center of Science, Tokyo Institute of
Technology, Tokyo 152-8551, Japan}
\author{Hiroyuki Hyuga}
\affiliation{Department of Physics, Keio University, Yokohama 223-8522,
Japan}

\date{\today}

\begin{abstract}
The moving-mirror problem is microscopically formulated without invoking
the external boundary conditions.
The moving mirrors are described by the quantized matter field
interacting with the photon field, forming dynamical cavity polaritons:
photons in the cavity are dressed by electrons in the moving mirrors.
The effective Hamiltonian for the polariton is derived, and corrections to 
the results based on the external boundary conditions are discussed.
\end{abstract}


\maketitle

\section{Introduction}
When mirrors are closely placed, the attractive or repulsive force between
them is observed.
This phenomenon, known as the Casimir effect~\cite{Casimir}, is explained
by the fact that the vacuum state of the electromagnetic (EM) field in the
presence of the mirrors is modified from that of the free space, and the
vacuum fluctuation energy depends on the positions of the mirrors.
On the other hand, when the mirrors move very rapidly, quantum state of
the EM field cannot adiabatically follow the instantaneous vacuum state
for each position of the mirrors, resulting in the creation of photons.
Such excitation of the quantum field caused by non-adiabatic change of
the vacuum state~\cite{Parker,Moore,Fulling} is referred to as the
dynamical Casimir effect (DCE), and there have been numerous
investigations into this
subject~\cite{Castagnino,Dodonov89,Jaekel,Neto,Dodonov90,Sarkar,Razavy,Law,Law95,Schutz,Yab,Dodonov93,Okushima,Saito,Barton,Salamone,Barton95,Barton96,Gutig,Eberlein99,Dodonov98,Lambrecht,Plunien,Schwinger,Eberlein,Dodonov96,Dodonov01},
e.g., spectral properties of created photons~\cite{Castagnino},
radiation pressure on a moving mirror~\cite{Dodonov89,Jaekel,Neto},
squeezing in the radiation field~\cite{Dodonov90,Sarkar},
effective Hamiltonian approach~\cite{Razavy,Law,Law95,Schutz},
time-varying refractive index~\cite{Yab,Dodonov93,Okushima,Saito},
radiation from moving
dielectrics~\cite{Barton,Salamone,Barton95,Barton96,Gutig,Eberlein99},
influence of finite temperature~\cite{Dodonov98,Lambrecht,Plunien},
and relation with sonoluminescence~\cite{Schwinger,Eberlein}.

In the most of the previous works, the moving mirrors have been treated as 
the moving boundary conditions such that the transverse components of the
electric field operator vanish at the mirror surfaces in their rest
frames.
Such external boundary conditions, of course, violate the commutation
relation of the EM field operators at the boundaries, and the
incarceration of the photon field between the moving mirrors causes the
temporal change of the Hilbert space.
Thus, the `classical' external boundary conditions involve quantum
mechanical imperfections.
To circumvent these conceptual difficulties, the various results of the
moving-mirror problem based on the boundary conditions should be examined
and derived as some limiting case of more elaborate models.
Several studies have been done towards this direction by considering
moving matter with finite refractive
index~\cite{Barton,Salamone,Barton95,Barton96,Gutig,Eberlein99}:
formulation of the problem and radiation spectrum~\cite{Barton},
radiative reaction on the mirror~\cite{Barton,Salamone},
the dispersive mirror~\cite{Salamone,Barton95},
radiation in two and three dimension~\cite{Barton96,Gutig},
and density variation in dielectrics~\cite{Eberlein99}.

Recently, Koashi and Ueda~\cite{Koashi} formulated the static Casimir
effect based on a combined system of the EM and matter fields, and
showed that the both fields participate in the vacuum fluctuations
inducing the Casimir force.
Although the quantum theory of the systems in which the EM field and
matter interact with each other has been developed by many
authors~\cite{Fano,Hopfield,Glauber,Huttner,Matloob,Koashi}, such an
approach to the DCE has hardly been made so far~\cite{Okushima}.

The aim of the present paper is to formulate the DCE in a moving-matter
system in terms of the quantized field-matter theory.
The EM field attenuates by coupling with the matter field inside the
mirrors, and therefore no external boundary conditions are required.
In other words, the EM field confined in the resonator is dressed by the
matter field inside the mirrors, forming cavity polaritons.
Non-adiabatic movement of the mirrors excites the cavity polaritons and
this phenomenon may be called the DCE of polaritons.
In this paper, we derive the effective Hamiltonian for polaritons, apply
it to the moving-mirror problem, and compare the result with that based on
the external boundary conditions~\cite{Law}.

This paper is organized as follows.
In Sec.~\ref{s:review}, the moving-mirror problem is briefly reviewed.
In Sec.~\ref{s:formalism}, we formulate quantum theory of field-matter
interacting systems, in which matter is allowed to move.
In Sec.~\ref{s:Heff}, we derive the effective Hamiltonian for polaritons
in the moving-mirror system, and apply it to the one-dimensional case.
Final section presents the summary of this paper, and some complicated
algebraic manipulations are relegated to appendices.

\section{Brief review of the moving-mirror problem}
\label{s:review}

\begin{figure}[tb]
\includegraphics[width=8.6cm]{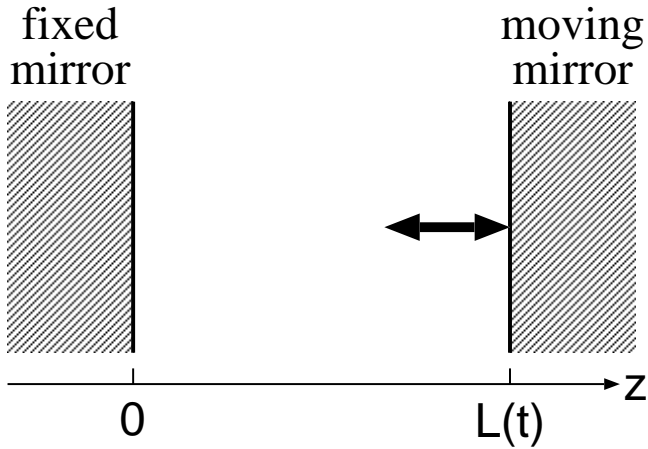}
\caption{
Schematic illustration of the one-dimensional moving-mirror problem.
The left mirror is fixed at $z = 0$, and the right mirror moves along the
$z$ axis.
}
\label{f:mirror}
\end{figure}

We briefly review the moving-mirror problem~\cite{Moore} to make the
present paper self-contained and to fix the notation.
The simplest system consists of two perfectly reflecting mirror plates
placed in parallel as illustrated in
Fig.~\ref{f:mirror}.
The mirror at the origin $z = 0$ is fixed and the other at the position $z
= L(t)$ is allowed to move.
The system is assumed to be uniform in the $x$ and $y$ directions, and
we consider only one component of the vector potential, say the $x$
component $A_x(z, t)$, without loss of generality.
The vector potential in the Coulomb gauge obeys the wave equation as (we
omit the subscript $x$ of $A_x$ from now on)
\begin{equation} \label{waveeq}
\frac{\partial^2 A(z, t)}{c^2 \partial t^2}
= \frac{\partial^2 A(z, t)}{\partial z^2},
\end{equation}
and the boundary conditions are imposed as $A(0, t) = A(L(t), t) = 0$,
which guarantee that the transverse components of the electric field
vanish at the surfaces of the mirrors in their rest frames.
The field operator of the vector potential $\hat A(z, t)$ in the
Heisenberg representation can be expanded as
\begin{equation} \label{Ax}
\hat A(z, t) = \sum_n \sqrt{\frac{\hbar}{2 \varepsilon_0}}
\left[ \hat a_n(t) f_n(z, t) + \hat a_n^\dagger(t) f_n^*(z, t) \right],
\end{equation}
where $\hat a_n$ and $\hat a_n^\dagger$ are the annihilation and creation
operators of photons of the $n$th mode.

One approach~\cite{Moore} to this problem is to fix $\hat a_n$ and $\hat
a_n^\dagger$ to ones at $t = 0$, and evolve the function $f_n(z, t)$ as
\begin{equation} \label{waveeq2}
\frac{\partial^2 f_n(z, t)}{c^2 \partial t^2} = \frac{\partial^2
f_n(z, t)}{\partial z^2}
\end{equation}
with the boundary conditions
\begin{equation} \label{bc}
f_n(0, t) = f_n(L(t), t) = 0,
\end{equation}
which ensure that $\hat A(z, t)$ obeys the wave equation (\ref{waveeq})
and the boundary conditions $\hat A(0, t) = \hat A(L(t), t) = 0$.
When $L$ is constant, the function $f_n(z, t)$ is given by
\begin{equation}
f_n(z, t) = \sqrt{\frac{2}{\omega_n L}} e^{-i \omega_n t} \sin k_n z,
\end{equation}
where $k_n = n \pi / L$ and $\omega_n = c k_n$.
When the characteristic time of the mirror motion is much larger than
$L(t) / c$, $f_n(z, t)$  adiabatically follows the mode function for each
$L(t)$ as
\begin{equation} \label{adf}
f_n(z, t) \simeq \sqrt{\frac{2}{\omega_n(t) L(t)}} e^{-i \int_0^t
\omega_n(\tau) d\tau} \sin k_n(t) z,
\end{equation}
where $k_n(t) = n \pi / L(t)$ and $\omega_n(t) = c k_n(t)$.
When the mirror moves much faster, the adiabatic theorem breaks down, and
$f_n(z, t)$ evolves in a more complicated manner.
The time evolution of the system in this approach is thus not generated by
a predetermined Hamiltonian but by the classical equation of motion
(\ref{waveeq2}) and the boundary conditions (\ref{bc}), by which the time
evolution of $\hat A(z, t)$ in the Heisenberg representation is obtained.

Another approach to the moving-mirror problem is the method of an
effective Hamiltonian~\cite{Razavy,Law,Schutz}, where the function $f_n(z,
t)$ is fixed to the mode function for each $L(t)$ as
\begin{equation}
f_n(z, t) = \sqrt{\frac{2}{\omega_n(t) L(t)}} \sin k_n(t) z,
\end{equation}
and $\hat a_n(t)$ and $\hat a_n^\dagger(t)$ are time dependent.
This time evolution is described by the Heisenberg equation as
\begin{equation}
i \hbar \frac{d \hat a(t)}{dt} = [ \hat a(t), \hat H_a^{\rm eff}(t)],
\end{equation}
where the effective Hamiltonian for an operator
$\hat {\cal O}$ is denoted by $\hat H_{\cal O}^{\rm eff}(t)$.
In the one-dimensional moving-mirror problem, the effective Hamiltonian
for $\hat a_n$ and $\hat a_n^\dagger$ is obtained as~\cite{Law} (see
Appendix~\ref{app:Law} for the derivation)
\begin{widetext}
\begin{equation} \label{HEFFLAW}
\frac{\hat H_a^{\rm eff}(t)}{\hbar} = \sum_n \omega_n(t) \hat
a_n^\dagger \hat a_n - \frac{i \dot L(t)}{4 L(t)} \sum_n (\hat
a_n^{\dagger 2} - \hat a_n^2) + \frac{i \dot L(t)}{L(t)} \sum_{n \neq n'}
(-1)^{n + n'} \frac{nn'}{n^2 - n^{\prime 2}} \sqrt{\frac{n'}{n}} (\hat
a_n^\dagger + \hat a_n) (\hat a_{n'}^\dagger - \hat a_{n'}).
\end{equation}
\end{widetext}
If the first term is dominant on the right-hand side of
Eq.~(\ref{HEFFLAW}), i.e., $\dot L / L \ll \omega_n$ for any $n$, the time
evolution operator approximately becomes $\exp\left[-i \int_0^t d\tau
\sum_n \omega_n(\tau) \hat a_n^\dagger \hat a_n \right]$, showing the
adiabatic theorem.

One of the advantages of the effective Hamiltonian approach is that we
can study the state evolution in the Schr\"odinger representation by
\begin{equation}
i \hbar \frac{\partial}{\partial t} |\psi(t) \rangle = \hat H_a^{\rm
eff}(t) |\psi(t) \rangle.
\end{equation}
In this case, we note that the operators $\hat a_n$ and $\hat a_n^\dagger$
can be interpreted as the annihilation and creation operators of photons
of the $n$th mode defined in the interval $0 \leq z \leq L(t)$.
Another advantage of the effective Hamiltonian approach is that we can
understand easily what kinds of elementary processes occur.
For example, the second term of the effective Hamiltonian (\ref{HEFFLAW})
has the form of the parametric process, suggesting that the squeezed state
is produced when the mirror oscillates at frequency $2 \omega_n$.
The third term induces pair creation and annihilation of photons and
energy transfer between different modes.

In the above approaches, the boundary conditions that the filed operator
$\hat A(z, t)$ vanishes at $z = 0$ and $z = L(t)$ are imposed.
Because of these boundary conditions, the canonical commutation relation
$[\hat {\bf A}({\bf r}, t), \hat {\bf E}({\bf r}', t)] = -i \hbar /
\varepsilon_0 \mbox{\boldmath$\delta$}_T({\bf r} - {\bf r}')$ does not
hold at the boundaries, where $\mbox{\boldmath$\delta$}_T$ is the
transverse delta function.
The field operator (\ref{Ax}) and the Hilbert space on which it operates
are defined only within the interval $0 \leq z \leq L(t)$, and then the
Hilbert space varies accordingly.

\section{Formulation of the field-matter interacting systems}
\label{s:formalism}

\subsection{Field representation of systems}

We start from a classical microscopic model, in which polarizable atoms in
the matter interact with the EM field.
We suppose that the $i$th atom consists of an electron with charge $-e$
and mass $m_e$ at the position ${\bf r}_i$ and an ion with charge $+e$ and
mass $M$ at the position ${\bf R}_i$.
The relative vector between the electron and the ion is denoted by ${\bf
x}_i \equiv {\bf r}_i - {\bf R}_i$, and the center-of-mass vector is
$\bm{\Xi}_i \equiv (M {\bf R}_i + m_e {\bf r}_i) / (M + m_e) \simeq {\bf
R}_i$.
We assume that the center-of-mass vectors $\bm{\Xi}_i(t)$ are given
functions of time when the matter is moved.
The kinetic energy of the $i$th atoms is expressed as $m_e \dot{\bf r}_i^2
/ 2 + M \dot{\bf R}_i^2 / 2 = (M + m_e) \dot{\bm{\Xi}}_i^2 / 2 + m
\dot{\bf x}_i^2 / 2$, where $(M + m_e) \dot{\bm{\Xi}}_i^2 / 2$ is a known
function of time.
Therefore, we consider $m \dot{\bf x}_i^2 / 2$ alone in the dynamics,
where $m \equiv M m_e / (M + m_e) \simeq m_e$ is the reduced mass.
Furthermore the electrons and ions are assumed to be bounded by the
effective potential $m \Omega^2 {\bf x}_i^2 / 2$.
The Lagrangian for this system is then given by
\begin{eqnarray} \label{L}
L & = & \int d{\bf r} \left[ \frac{\varepsilon_0}{2} {\bf E}^2({\bf r},
t) - \frac{1}{2\mu_0} {\bf B}^2({\bf r}, t) \right] \nonumber \\
& & + \sum_i \left[ \frac{m}{2} \dot{\bf x}_i^2 - \frac{m \Omega^2}{2}
{\bf x}_i^2 \right] - e \sum_i \left[ \phi({\bf R}_i, t) - \phi({\bf r}_i,
t) \right] \nonumber \\
& & + e \sum_i \left[ \dot{\bf R}_i \cdot {\bf A}({\bf R}_i, t)
- \dot{\bf r}_i \cdot {\bf A}({\bf r}_i, t) \right],
\end{eqnarray}
where ${\bf E} = -\nabla \phi - \partial {\bf A} / \partial t$ and ${\bf
B} = \nabla \times {\bf A}$.
The Euler-Lagrange equations are obtained as
\begin{subequations}
\begin{eqnarray}
& & \varepsilon_0 \nabla \cdot {\bf E}({\bf r}, t) - e \sum_i
\left[ \delta({\bf r} - {\bf R}_i) - \delta({\bf r} - {\bf r}_i)
\right] = 0, \nonumber \\
& & \\
& & \varepsilon_0 \frac{\partial {\bf E}({\bf r}, t)}{\partial t} -
\frac{1}{\mu_0} \nabla \times {\bf B}({\bf r}, t) \nonumber \\
& & + e \sum_i \left[ \dot{\bf R}_i \delta({\bf r} - {\bf R}_i) - \dot{\bf
r}_i \delta({\bf r} - {\bf r}_i) \right] = 0, \\
& & \ddot{\bf x}_i = -\Omega^2 {\bf x}_i - \frac{e}{m_e} \left[ {\bf
E}({\bf r}_i, t) + \dot{\bf r}_i \times {\bf B}({\bf r}_i, t) \right]
\nonumber \\
& & - \frac{e}{M} \left[ {\bf E}({\bf R}_i, t) + \dot{\bf R}_i \times {\bf
B}({\bf R}_i, t) \right].
\end{eqnarray}
\end{subequations}
The first and second equations are the Maxwell equations, and the third
one describes the motion of charged particles in the EM field.

We rewrite the above particle picture of the polarizable atoms in terms
of the field picture.
When the difference between adjacent polarizations $|{\bf x}_{i + 1} -
{\bf x}_i|$ is much smaller than the characteristic amplitudes of $|{\bf
x}_i|$ and $|{\bf x}_{i + 1}|$, namely the polarizations change smoothly
in the lattice scale, we can replace ${\bf x}_i$ with the polarization
field ${\bf X}$ as
\begin{equation} \label{ptof1}
{\bf x}_i(t) \Longrightarrow {\bf X}({\bf r}, t).
\end{equation}
The density of the polarizable atoms is replaced as
\begin{equation} \label{ptof2}
\sum_i \delta({\bf r} - \bm{\Xi}_i(t)) \Longrightarrow \rho({\bf r}, t).
\end{equation}
The polarization field ${\bf X}$ and the density $\rho$ vanish outside the 
matter.
The velocity of the matter $\dot{\bm{\Xi}}_i(t)$ is denoted by ${\bf
v}({\bf r}, t)$, which is defined only inside the matter.
The time derivative $\dot{\bf x}_i(t)$ should be replaced by $d {\bf
X}({\bf r}, t) / dt$ with
\begin{equation}
\frac{d}{dt} \equiv \frac{\partial}{\partial t} + {\bf v}({\bf r}, t)
\cdot \nabla.
\end{equation}
In the present paper, for simplicity, we consider the case in which the
matter is allowed to undergo only translational motion.
Rotations and deformations of the matter complicate extremely the
formulation and are not considered.
In this case, the velocity of the matter is uniform in each object, i.e.,
$\nabla {\bf v}({\bf r}, t) = 0$.
For example, in the case of Fig.~\ref{f:mirror}, $v(z, t) = 0$ for $z \leq
0$ and $v(z, t) = \dot L(t)$ for $z \geq L(t)$.
The density $\rho({\bf r}, t)$ becomes a function of ${\bf r} - {\bf v} t$
with ${\bf v}$ the velocity of each object, giving
\begin{equation}
\frac{d \rho({\bf r}, t)}{dt} = \left( \frac{\partial}{\partial
 t} + {\bf v} \cdot \nabla \right) \rho({\bf r}, t) = 0.
\end{equation}

Using Eqs.~(\ref{ptof1}) and (\ref{ptof2}), the first summation in the
Lagrangian (\ref{L}) is replaced by
\begin{equation} \label{matter}
\frac{m}{2} \int d{\bf r} \rho({\bf r}, t) \left[ \left( \frac{d {\bf
X}({\bf r}, t)}{dt} \right)^2 - \Omega^2 {\bf X}^2({\bf r}, t) \right].
\end{equation}
The interaction terms in the Lagrangian (\ref{L}) can be rewritten as
\begin{eqnarray} \label{int}
& & \sum_i \left[ \phi({\bf R}_i, t) - \phi({\bf r}_i, t) - \dot{\bf R}_i
\cdot {\bf A}({\bf R_i}, t) + \dot{\bf r}_i \cdot {\bf A}({\bf r}_i, t)
\right]
\nonumber \\
& = & \sum_i \int d{\bf r} \Biggl[ \phi\left( {\bf r} - \frac{m}{M} {\bf
x}_i, t \right) - \phi\left( {\bf r} + \frac{m}{m_e} {\bf x}_i, t \right)
\nonumber \\
& & - \dot{\bf R}_i \cdot {\bf A}\left( {\bf r} - \frac{m}{M} {\bf x}_i, t
\right) + \dot{\bf r}_i \cdot {\bf A}\left( {\bf r} + \frac{m}{m_e} {\bf
x}_i, t \right) \Biggr] \nonumber \\
& & \times \delta({\bf r} - \bm{\Xi}_i) \nonumber \\
& \simeq & \sum_i \int d{\bf r} \biggl[ -{\bf x}_i \cdot \nabla \phi({\bf
r}, t) + \dot{\bf x}_i \cdot {\bf A}({\bf r}, t) \nonumber \\
& & + \dot{\bm{\Xi}}_i \cdot [{\bf x}_i \cdot \nabla] {\bf A}({\bf r}, t)
\biggr] \delta({\bf r} - \bm{\Xi}_i),
\end{eqnarray}
where in the last line we assumed that $\phi$ and ${\bf A}$ are slowly
varying functions in the scale of $|{\bf x}_i|$, and ignored the second
and higher order of $|{\bf x}_i|$.
Applying the replacements (\ref{ptof1}) and (\ref{ptof2}) to
Eq.~(\ref{int}) yields
\begin{eqnarray} \label{int2}
& & \int d{\bf r} \rho({\bf r}, t) \biggl\{ -{\bf X}({\bf r}, t) \cdot
\nabla \phi({\bf r}, t) + \frac{d {\bf X}({\bf r}, t)}{dt} \cdot {\bf
A}({\bf r}, t) \nonumber \\
& & + {\bf v}({\bf r}, t) \cdot [{\bf X} ({\bf r}, t) \cdot 
\nabla] {\bf A}({\bf r}, t) \biggr\} \nonumber \\
& = & \int d{\bf r} \rho({\bf r}, t) {\bf X}({\bf r}, t) \cdot
[{\bf E}({\bf r}, t) + {\bf v}({\bf r}, t) \times {\bf B}({\bf r}, t)] +
\nonumber \\
& & \frac{d}{dt} \int d{\bf r} \rho({\bf r}, t) {\bf X} ({\bf r}, t) \cdot
{\bf A}({\bf r}, t),
\end{eqnarray}
where the second integral in the second line can be ignored because the
total derivative term in the Lagrangian is irrelevant in the dynamics.
Thus, from Eqs.~(\ref{matter}) and (\ref{int2}), the Lagrangian for the
system is obtained as
\begin{eqnarray} \label{Lag}
L & = & \int d{\bf r} \Biggl\{ \frac{\varepsilon_0}{2} {\bf E}^2({\bf r},
t) - \frac{1}{2\mu_0} {\bf B}^2({\bf r}, t) \nonumber \\
& & + \frac{m}{2} \rho({\bf r}, t) \left[ \left( \frac{d {\bf X}({\bf r},
t)}{dt} \right)^2 - \Omega^2 {\bf X}^2({\bf r}, t) \right]
\nonumber \\
& & - e \rho({\bf r}, t) {\bf X}({\bf r}, t) \cdot [{\bf E}({\bf r}, t) +
{\bf v}({\bf r}, t) \times {\bf B}({\bf r}, t)] \Biggl\}.
\end{eqnarray}
We note that this Lagrangian reduces to the one used in the static and
uniform dielectrics~\cite{Huttner}, when the matter is fixed and the
density $\rho$ is uniform.

The conjugate momenta of ${\bf A}$ and ${\bf X}$ are given by
\begin{eqnarray}
\label{Pi}
{\bf \Pi}({\bf r}, t) & \equiv & \frac{\delta L}{\delta [\partial_t {\bf
A}({\bf r}, t)]} \nonumber \\
& = & -\varepsilon_0 {\bf E}({\bf r}, t) + e \rho({\bf r}, t)
{\bf X}({\bf r}, t), \\
\label{Y}
{\bf Y}({\bf r}, t) & \equiv & \frac{\delta L}{\delta [\partial_t
{\bf X}({\bf r}, t)]} =  m \rho({\bf r}, t) \frac{d {\bf X}({\bf r},
t)}{dt}.
\end{eqnarray}
The Euler-Lagrange equation for $\phi$ reads
\begin{equation} \label{eom1}
\nabla \cdot {\bf D}({\bf r}, t) = 0,
\end{equation}
where ${\bf D} \equiv -{\bf \Pi}$ can be regarded as the electric
displacement with polarization $-e \rho {\bf X}$.
Adopting the Coulomb gauge ($\nabla \cdot {\bf A} = 0$) in
Eq.~(\ref{eom1}), we can write the electric potential $\phi$ as
\begin{equation} \label{phi}
\phi({\bf r}, t) = -\frac{e}{\varepsilon_0} \frac{1}{\nabla^2} \nabla
\cdot \left[ \rho({\bf r}, t) {\bf X}({\bf r}, t) \right],
\end{equation}
where $\frac{1}{\nabla^2} f({\bf r}) \equiv -\int d{\bf r}' f({\bf r}') /
(4 \pi |{\bf r} - {\bf r}'|)$.
The Euler-Lagrange equations for ${\bf A}$ and ${\bf X}$ are obtained as
\begin{eqnarray}
\label{eom2}
\varepsilon_0 \frac{\partial {\bf E}({\bf r}, t)}{\partial t} & = &
\frac{1}{\mu_0} \nabla \times {\bf B}({\bf r}, t) + e \rho({\bf r}, t)
\frac{d {\bf X}({\bf r}, t)}{dt} \nonumber \\
& & - e {\bf v}({\bf r}, t) \nabla \cdot
[\rho({\bf r}, t) {\bf X}({\bf r}, t)], \\
\label{eom3}
m \frac{d^2 {\bf X}({\bf r}, t)}{d t^2} & = & -m \Omega^2 {\bf X}({\bf r},
t) \nonumber \\
& & - e [{\bf E}({\bf r}, t) + {\bf v}({\bf r}, t) \times {\bf B}({\bf r},
t)].
\end{eqnarray}
Equation (\ref{eom2}) corresponds to the Maxwell equation $\nabla \times
{\bf B} / \mu_0 = {\bf J} + \partial {\bf D} / \partial t$ with current
${\bf J} \equiv e {\bf v} \nabla \cdot (\rho {\bf X}) - e ({\bf v} \cdot
\nabla) \rho {\bf X}$, which satisfies $\nabla \cdot {\bf J} = 0$.
Equation (\ref{eom3}) describes the dynamics of polarization moving in the
EM field.
Using Eqs.~(\ref{Lag})-(\ref{eom1}), we obtain the Hamiltonian as
\begin{eqnarray} \label{H}
H & = & \int d{\bf r} \left[ {\bf\Pi}({\bf r}, t) \cdot \frac{\partial
{\bf A}({\bf r}, t)}{\partial t} + {\bf Y}({\bf r}, t) \cdot
\frac{\partial {\bf X}({\bf r}, t)}{\partial t} \right] - L \nonumber \\
& = & \int d{\bf r} \Biggl\{ \frac{1}{2\varepsilon_0} \left[
{\bf\Pi}({\bf r}, t) - e \rho({\bf r}, t) {\bf X}({\bf r}, t)
\right]^2 + \frac{1}{2\mu_0} {\bf B}^2({\bf r}, t) \nonumber \\
& & + \frac{1}{2 m \rho({\bf r}, t)} {\bf Y}^2({\bf r}, t) + \frac{m
\Omega^2}{2} \rho({\bf r}, t) {\bf X}^2({\bf r}, t) \nonumber \\
& & + e \rho({\bf r}, t) {\bf X}({\bf r}, t) \cdot [{\bf v}({\bf r}, t)
\times {\bf B}({\bf r}, t)] \nonumber \\
& & - {\bf Y}({\bf r}, t) \cdot [{\bf v}({\bf r},
t) \cdot \nabla] {\bf X}({\bf r}, t) \Biggr\}.
\end{eqnarray}
Equations (\ref{Pi}), (\ref{Y}), (\ref{eom2}), and (\ref{eom3}) are
derived as the canonical equations of this Hamiltonian.

We note that the Hamiltonian (\ref{H}) reduces to the macroscopic model
described in terms of the dielectric constant
$\varepsilon$~\cite{Barton,Salamone,Barton95,Barton96,Gutig,Eberlein99,Eberlein},
when the dynamics of polarization can be eliminated.
The Hamiltonian in this model is given by
\begin{eqnarray} \label{macroH}
H & = & \int d{\bf r} \frac{1}{2} \left( {\bf E} \cdot {\bf D} + {\bf B}
\cdot {\bf H} \right) \nonumber \\
& = & \int d{\bf r} \frac{1}{2} \left[ {\bf E} \cdot {\bf D} +
\frac{1}{\mu_0} {\bf B}^2 + \left( 1 - \frac{1}{\varepsilon} \right)
{\bf v} \cdot ({\bf D} \times {\bf B}) \right] \nonumber \\
& & + O(v^2),
\end{eqnarray}
where we used the relations for moving medium~\cite{Landau}
\begin{subequations}
\begin{eqnarray}
{\bf D} & = & \varepsilon {\bf E} + (\varepsilon - \varepsilon_0) {\bf v}
\times {\bf B} + O(v^2), \label{DE} \\
{\bf B} & = & \mu_0 {\bf H} + (\varepsilon - \varepsilon_0) {\bf E} \times
{\bf v} + O(v^2).
\end{eqnarray}
\end{subequations}
If we neglect the left-hand side of Eq.~(\ref{eom3}), we get
\begin{equation} \label{staticX}
{\bf X} = -\frac{e}{m \Omega^2} ({\bf E} + {\bf v} \times {\bf B}).
\end{equation}
This relation and ${\bf D} = \varepsilon_0 {\bf E} - e \rho {\bf X}$ yield
Eq.~(\ref{DE}) with $\varepsilon = 1 + e^2 \rho / \varepsilon_0 m
\Omega^2$.
Substituting Eq.~(\ref{staticX}) into our Hamiltonian (\ref{H}), and
dropping the terms including ${\bf Y}$, we obtain Eq.~(\ref{macroH}).
Our model, therefore, reduces to the above dielectric model, when the
dynamics of polarization is neglected.
This corresponds to the case in which the relevant frequencies of the EM
field are much smaller than the characteristic frequencies of matter
$\Omega$ and $\omega_p$.

\subsection{Quantization: polaritons}

We find out the normal mode of the field-matter coupled equations, in
which the positions of the matter are fixed, i.e., ${\bf v} = 0$.
We denote this fixed matter configuration as $\cal M$, and the mode
functions and frequencies depend on it: $A_n({\bf r}, {\cal M})$,
$\cdots$, $\omega_n({\cal M})$, where $n$ is the index of the mode.
For brevity, we omit the argument ${\cal M}$ below.
Substitution of $(\hbar / 2\omega_n)^{1/2} {\bf A}_n({\bf r}) e^{-i
\omega_n t}$, $i (\hbar \omega_n / 2)^{1/2} {\bf \Pi}_n({\bf r}) e^{-i
\omega_n t}$, $i (\hbar \omega_n / 2)^{1/2} {\bf X}_n({\bf r}) e^{-i
\omega_n t}$, and $(\hbar / 2\omega_n)^{1/2} {\bf Y}_n({\bf r}) e^{-i
\omega_n t}$ into ${\bf A}$, ${\bf \Pi}$, ${\bf X}$, and ${\bf Y}$ in
Eqs.~(\ref{Pi}), (\ref{Y}), (\ref{eom2}), and (\ref{eom3}) for ${\bf v} =
0$ yields the normal-mode equations
\begin{subequations} \label{mode}
\begin{eqnarray}
{\bf \Pi}_n({\bf r}) & = & -\varepsilon_0 {\bf A}_n({\bf r}) \nonumber \\
& & + e \left( 1
- \nabla \frac{1}{\nabla^2} \nabla \cdot \right) [\rho({\bf r}) {\bf
X}_n({\bf r})], \\
\omega_n^2 {\bf \Pi}_n({\bf r}) & = & \frac{1}{\mu_0} \nabla^2 {\bf
A}_n({\bf r}), \\
{\bf Y}_n({\bf r}) & = & m \omega_n^2 \rho({\bf r}) {\bf X}_n({\bf r}), \\
{\bf Y}_n({\bf r}) & = & \left[ m \Omega^2 \rho({\bf r}) +
\frac{e^2}{\varepsilon_0} \rho^2({\bf r}) \right] {\bf X}_n({\bf r})
\nonumber \\
& & - \frac{e}{\varepsilon_0} \rho({\bf r}) {\bf \Pi}_n({\bf r}),
\end{eqnarray}
\end{subequations}
where we used Eq.~(\ref{phi}).
We take ${\bf A}_n$, ${\bf \Pi}_n$, ${\bf X}_n$, and ${\bf Y}_n$ to be
real without loss of generality.
These mode functions can be shown to satisfy the orthonormal relation
(see Appendix~\ref{app:orth})
\begin{equation} \label{orthonormal}
\int d{\bf r} \left[ {\bf A}_n({\bf r}) \cdot {\bf \Pi}_{n'}({\bf r}) -
{\bf Y}_n({\bf r}) \cdot {\bf X}_{n'}({\bf r}) \right] = -\delta_{nn'}.
\end{equation}

In terms of the mode functions, we can expand the time evolution of the
fields as
\begin{subequations} \label{expand}
\begin{eqnarray}
{\bf A}({\bf r}, t) & = & \sum_n \sqrt{\frac{\hbar}{2\omega_n}} {\bf
A}_n({\bf r}) \left( b_n e^{-i \omega_n t} + {\rm c.c.} \right), \\
{\bf \Pi}({\bf r}, t) & = & \sum_n i \sqrt{\frac{\hbar \omega_n}{2}} {\bf
\Pi}_n({\bf r}) \left( b_n e^{-i \omega_n t} - {\rm c.c.} \right), \\
{\bf X}({\bf r}, t) & = & \sum_n i \sqrt{\frac{\hbar \omega_n}{2}}
{\bf X}_n({\bf r}) \left( b_n e^{-i \omega_n t} - {\rm c.c.}
\right), \\
{\bf Y}({\bf r}, t) & = & \sum_n \sqrt{\frac{\hbar}{2\omega_n}} {\bf
Y}_n({\bf r}) \left( b_n e^{-i \omega_n t} + {\rm c.c.} \right),
\end{eqnarray}
\end{subequations}
where $b_n$ is the complex amplitude of each mode, and ${\rm c.c.}$
indicates the complex conjugate of the previous term.
Substituting the expansions (\ref{expand}) into the Hamiltonian (\ref{H})
for ${\bf v} = 0$, denoted by $H_{v = 0}$, and using Eqs.~(\ref{mode}), we
obtain
\begin{equation}
H_{v = 0} = \sum_n \hbar \omega_n b_n^* b_n.
\end{equation}
Following the standard quantization procedure, we replace the c-numbers
$b_n$ and $b_n^*$ with the Bose operators $\hat b_n$ and $\hat
b_n^\dagger$ satisfying the commutation relation $[\hat b_n, \hat
b_{n'}^\dagger] = \delta_{nn'}$.
The elementary excitations created by $\hat b_n^\dagger$ can be regarded
as polaritons, since they are linear combinations of the photon and
polarization fields.
We should note that the operators $\hat b_n$ also depend on the matter
configuration $\cal M$, and are to be expressed as $\hat b_n({\cal M})$ in
full detail.
The field operators in the Schr\"odinger representation can be expanded as 
\begin{subequations} \label{opexpand}
\begin{eqnarray}
\hat {\bf A}({\bf r}) & = & \sum_n \sqrt{\frac{\hbar}{2\omega_n}}
{\bf A}_n({\bf r}) (\hat b_n + \hat b_n^\dagger), \\
\hat {\bf \Pi}({\bf r}) & = & \sum_n i \sqrt{\frac{\hbar \omega_n}{2}}
{\bf \Pi}_n({\bf r}) (\hat b_n - \hat b_n^\dagger), \\
\hat {\bf X}({\bf r}) & = & \sum_n i \sqrt{\frac{\hbar \omega_n}{2}} 
{\bf X}_n({\bf r}) (\hat b_n - \hat b_n^\dagger), \\
\hat {\bf Y}({\bf r}) & = & \sum_n \sqrt{\frac{\hbar}{2\omega_n}}
{\bf Y}_n({\bf r}) (\hat b_n + \hat b_n^\dagger).
\end{eqnarray}
\end{subequations}

In the above argument, we assumed the discrete spectrum of polaritons.
When the spectrum is continuous, the continuous index of the mode, such as 
wave number $k$, should be used instead of $n$, and the summation $\sum_n$ 
should be replaced by an appropriate integral.

\section{Field-matter formalism of the moving-mirror problem}
\label{s:Heff}

\subsection{The effective Hamiltonian}

In the previous section, the polaritons were derived for fixed matter
configuration $\cal M$.
The number states of the polaritons are eigenstates of the Hamiltonian for 
the system with $\cal M$, and thus suitable for orthogonal set of
bases in the Fock space of polaritons.
When the matter configuration transforms to ${\cal M}'$, definition of 
polaritons alters accordingly, and the number states of the polaritons in
${\cal M}'$ should be used as new bases.
As a result of change of bases, the state vector undergoes unitary
transformation.
Thus, when the matter configuration continuously transforms as ${\cal
M}(t)$ and we insist on using the Fock state bases in instantaneous matter
configuration, the state vector undergoes extra evolution in addition to
the usual time evolution.
This representation (the bases follow the eigenstates of the
time-dependent Hamiltonian instantaneously) is often employed in the
adiabatic approximation~\cite{Messiah}.
The effective Hamiltonian describing such state evolution can be derived
by a few ways~\cite{Razavy,Law,Schutz,Eberlein} that are equivalent each
other, and here we follow the one in Ref.~\cite{Eberlein}.

The Schr\"odinger equation is written by
\begin{eqnarray} \label{Sch}
i \hbar \frac{\partial}{\partial t} |\psi(t) \rangle & = & \hat H({\cal
M}(t)) |\psi(t) \rangle \nonumber \\
& = & [\hat H_{v = 0}({\cal M}(t)) + \hat K({\cal M}(t))] |\psi(t)
\rangle,
\end{eqnarray}
where $\hat H_{v = 0}$ is the part that does not include the velocity of
the matter explicitly, and $\hat K \equiv \hat H - \hat H_{v = 0}$.
Expanding the state vector as $|\psi(t) \rangle = \sum_i c_i(t)
|\psi_i({\cal M}(t)) \rangle$ with the eigenvectors satisfying
\begin{equation} \label{eigen}
[\hat H_{v = 0}({\cal M}(t)) - E_i({\cal M}(t))] |\psi_i({\cal M}(t))
\rangle = 0,
\end{equation}
Eq.~(\ref{Sch}) becomes
\begin{eqnarray}
i \hbar \dot c_i(t) & = & E_i c_i(t) + \sum_j \langle \psi_i
 | \hat K | \psi_j \rangle c_j(t) \nonumber \\
& & - i \hbar \dot{\cal M}(t) \sum_j \langle
\psi_i | \frac{\partial}{\partial {\cal M}} | \psi_j \rangle c_j(t)
\nonumber \\
& \equiv & \sum_j \langle \psi_i | \hat H^{\rm eff} | \psi_j \rangle
c_j(t),
\end{eqnarray}
where we omit the argument ${\cal M}(t)$ for brevity.
The symbol $\partial / \partial {\cal M}$ indicates differentiation
with respect to the positions of the matter, e.g., $\partial / \partial L$
in the situation of Fig.~\ref{f:mirror}.
The matrix element of the effective Hamiltonian is thus given by
\begin{eqnarray} \label{Heff0}
\langle \psi_i | \hat H^{\rm eff} | \psi_j \rangle & = & E_i \delta_{ij} +
\langle \psi_i | \hat K | \psi_j \rangle \nonumber \\
& & - i \hbar \dot{\cal M}(t) \langle
\psi_i | \frac{\partial}{\partial {\cal M}} | \psi_j \rangle.
\end{eqnarray}

Now we express the effective Hamiltonian by using the instantaneous
creation and annihilation operators $\hat b_n^\dagger({\cal M}(t))$ and
$\hat b_n({\cal M}(t))$ (we omit the argument ${\cal M}(t)$ below).
The first term on the right-hand side of Eq.~(\ref{Heff0}) corresponds to
$\sum_n \hbar \omega_n \hat b_n^\dagger \hat b_n$ in the effective
Hamiltonian.
Substituting the field expansions (\ref{opexpand}) into the
velocity-dependent part
\begin{eqnarray} \label{Hv}
\hat K & = & \int d{\bf r} \bigl\{ e \rho({\bf r}, t) \hat {\bf X}({\bf
r}, t) \cdot [{\bf v}({\bf r}, t) \times \hat {\bf B}({\bf r}, t)]
\nonumber \\
& & - \hat {\bf Y}({\bf r}, t) \cdot [{\bf v}({\bf r}, t) \cdot \nabla]
\hat {\bf X}({\bf r}, t) \bigr\},
\end{eqnarray}
we obtain
\begin{equation} \label{HeffK}
\hat K = i \hbar \sum_{n, n'} F_{nn'}^{(1)} (\hat b_n^\dagger + \hat
b_n)(\hat b_{n'}^\dagger - \hat b_{n'}),
\end{equation}
where we defined
\begin{eqnarray}
F^{(1)}_{nn'} & \equiv & -\frac{1}{2} \sqrt{\frac{\omega_{n'}}{\omega_n}}
\int d{\bf r} \Bigl\{ e \rho({\bf r}, t) \Bigl[ {\bf v}({\bf r}, t)
\cdot [{\bf X}_{n'}({\bf r}) \cdot \nabla] {\bf A}_n({\bf r}) \nonumber \\
& & - {\bf X}_{n'}({\bf r}) \cdot [{\bf v}({\bf r}, t) \cdot \nabla] {\bf
A}_n({\bf r}) \Bigr] \nonumber \\
& & - {\bf Y}_n({\bf r}) \cdot [{\bf v}({\bf r}, t) \cdot \nabla] {\bf 
X}_{n'}({\bf r}) \Bigr\}. \label{F1}
\end{eqnarray}

The last term on the right-hand side of Eq.~(\ref{Heff0}) is treated as
follows.
Differentiating the eigenvalue equation (\ref{eigen}) with respect to
${\cal M}$, we find
\begin{equation} \label{rar}
\langle \psi_i | \frac{\partial}{\partial {\cal M}} | \psi_j \rangle =
-\frac{\langle \psi_i | \frac{\partial \hat H_{v = 0}}{\partial {\cal M}}
| \psi_j \rangle}{E_i - E_j}
\end{equation}
for $i \neq j$, and we take the eigenstates as $\langle \psi_i |
\frac{\partial}{\partial {\cal M}} | \psi_i \rangle = 0$.
From Eqs.~(\ref{Heff0}) and (\ref{rar}), a term $\hat b_n \hat
b_{n'}^\dagger$ in $\frac{\partial \hat H_{v = 0}}{\partial {\cal M}}$,
for example, corresponds to $i \hat b_n \hat b_{n'}^\dagger / (\omega_{n'}
- \omega_n)$ in the effective Hamiltonian, since $E_i - E_j = \hbar
(\omega_{n'} - \omega_n)$ for non-vanishing matrix element.
The correspondences of terms in $\frac{\partial \hat H_{v = 0}}{\partial
{\cal M}}$ to those in the effective Hamiltonian are therefore obtained as
\begin{subequations} \label{subst}
\begin{eqnarray}
(\hat b_n + \hat b_n^\dagger)^2 & \rightarrow & \frac{i}{2 \omega_n}
(\hat b_n^{\dagger 2} - \hat b_n^2), \\
(\hat b_n \pm \hat b_n^\dagger)(\hat b_{n'} \pm \hat b_{n'}^\dagger) &
\rightarrow & \frac{i}{\omega_n + \omega_{n'}} (\hat b_n^\dagger \hat
b_{n'}^\dagger - \hat b_{n'} \hat b_n) \nonumber \\
& & \pm \frac{i}{\omega_n - \omega_{n'}} (\hat b_n^\dagger \hat b_{n'} -
\hat b_n b_{n'}^\dagger). \nonumber \\
& &
\end{eqnarray}
\end{subequations}
In the Hamiltonian (\ref{H}), the density $\rho$ depends on the matter
configuration, and then
\begin{widetext}
\begin{equation} \label{dHdt0}
\dot{\cal M}(t) \frac{\partial \hat H_{v = 0}}{\partial {\cal M}}  = 
\int d{\bf r} \frac{\partial \rho({\bf r}, t)}{\partial t} \Biggl\{
-\frac{e}{\varepsilon_0} \left[ \hat {\bf \Pi}({\bf r}, t) - e \rho({\bf
r}, t) \hat {\bf X}({\bf r}, t) \right] \cdot \hat {\bf X}({\bf r}, t)
- \frac{1}{2 m \rho^2({\bf r}, t)} \hat {\bf Y}^2({\bf r}, t) +
\frac{m \Omega^2}{2} \hat {\bf X}^2({\bf r}, t) \Biggr\}.
\end{equation}
Substituting the field expansions (\ref{opexpand}) into Eq.~(\ref{dHdt0}),
and applying Eqs.~(\ref{subst}), we find that the last term on the
right-hand side of Eq.~(\ref{Heff0}) becomes in the effective Hamiltonian
as
\begin{eqnarray} \label{HeffM}
& & i \hbar \sum_n \left( \frac{1}{2} F_{nn}^{(2)} -
\frac{1}{4} F_{nn}^{(2)} \right) \left( \hat b_n^{\dagger 2} - \hat b_n^2
\right) \nonumber \\
& & + i \hbar \sum_{n \neq n'} \sqrt{\omega_n \omega_{n'}} \Biggl[ \left(
F_{nn'}^{(2)} - \frac{1}{2} F_{nn'}^{(3)} \right) \frac{1}{\omega_n +
\omega_{n'}} (\hat b_n^\dagger \hat b_{n'}^\dagger - \hat b_{n'} \hat b_n)
- \left( F_{nn'}^{(2)} + \frac{1}{2} F_{nn'}^{(3)} \right)
\frac{1}{\omega_n - \omega_{n'}} (\hat b_n^\dagger \hat b_{n'} - \hat b_n
b_{n'}^\dagger) \Biggr],
\end{eqnarray}
where we defined
\begin{subequations} \label{F}
\begin{eqnarray}
F^{(2)}_{nn'} & \equiv & \int d{\bf r} \frac{\partial \rho({\bf r},
t)}{\partial t} \left\{ \frac{e}{2 \varepsilon_0} \left[ {\bf \Pi}_n({\bf
r}) - e \rho({\bf r}, t) {\bf X}_n({\bf r}) \right] \cdot {\bf
X}_{n'}({\bf r}) - \frac{m \Omega^2}{4} {\bf X}_n({\bf r}) \cdot {\bf
X}_{n'}({\bf r}) \right\}, \\
F^{(3)}_{nn'} & \equiv & \frac{1}{\omega_n \omega_{n'}} \int d{\bf r}
\frac{\partial \rho({\bf r}, t)}{\partial t} \frac{1}{2 m \rho^2({\bf r},
t)} {\bf Y}_n({\bf r}) \cdot {\bf Y}_{n'}({\bf r}).
\end{eqnarray}
\end{subequations}

Rearranging the terms in Eqs.~(\ref{HeffK}) and (\ref{HeffM}), we obtain
the effective Hamiltonian for polaritons as
\begin{equation} \label{HEFF}
\frac{\hat H_b^{\rm eff}(t)}{\hbar} = \sum_n \omega_n \hat b_n^\dagger
\hat b_n + i \sum_n C_{nn} \left( \hat b_n^{\dagger 2} - \hat b_n^2
\right) + i \sum_{n \neq n'} C_{nn'} (\hat b_n^\dagger + \hat b_n)(\hat
b_{n'}^\dagger - \hat b_{n'})
\end{equation}
with
\begin{equation} \label{coeff}
C_{nn'} \equiv \left\{ \begin{array}{ll} F^{(1)}_{nn} + \frac{1}{2}
F^{(2)}_{nn} - \frac{1}{4} F^{(3)}_{nn} & (n = n') \\
F^{(1)}_{nn'} + \frac{\sqrt{\omega_n \omega_{n'}}}{\omega_n^2 -
\omega_{n'}^2} \left[ \omega_n (F^{(2)}_{nn'} + F^{(2)}_{n'n}) +
\omega_{n'} F^{(3)}_{nn'} \right] & (n \neq n'). \end{array} \right.
\end{equation}
\end{widetext}
This effective Hamiltonian for the polaritons with moving matter is the
main result of the present paper.
It is interesting to note that the effective Hamiltonian for polaritons
(\ref{HEFF}) and that for photons based on the external boundary
conditions (\ref{HEFFLAW}) have a similar form with respect to the
creation and annihilation operators, suggesting that Eq.~(\ref{HEFF})
reduces to Eq.~(\ref{HEFFLAW}) in some limiting case.
This will be explicitly shown in Sec.~\ref{s:oneD} for one-dimensional
case.
It is also suggested that the squeezed state of polaritons will be
generated by oscillation of the matter at an appropriate frequency by
analogy with the case of photons~\cite{Dodonov90,Sarkar}.
The important difference between Eqs.~(\ref{HEFF}) and (\ref{HEFFLAW})
is that the time evolution is fully described in a common Hilbert space in 
Eq.~(\ref{HEFF}) in contrast to Eq.~(\ref{HEFFLAW}) in which the Hilbert
space changes by the mirror motion.

\subsection{One-dimensional case} \label{s:oneD}

In order to compare our effective Hamiltonian (\ref{HEFF}) with
Eq.~(\ref{HEFFLAW}), we consider the one-dimensional moving-mirror problem
as illustrated in Fig.~\ref{f:mirror}.
We assume that the system is uniform in the $x$ and $y$ directions, and
consider only the $x$ components of the vector fields $\hat A_x(z)$, $\hat
\Pi_x(z)$, $\hat X_x(z)$, and $\hat Y_x(z)$ without loss of generality (we
omit the subscript $x$ below).
The normal-mode equations (\ref{mode}) reduce to
\begin{subequations} \label{1Deqs}
\begin{eqnarray}
\Pi_n(z, t) & = & -\varepsilon_0 A_n(z, t) + e \rho(z, t) X(z, t), \\
\Pi_n(z, t) & = & \frac{1}{\mu_0 \omega_n^2(t)} A_n''(z, t), \\
Y_n(z, t) & = & m \omega_n^2(t) \rho(z, t) X_n(z, t), \\
Y_n(z, t) & = & \rho(z, t) \left[ m \Omega^2 X_n(z, t) + e A_n(z, t)
\right].
\end{eqnarray}
\end{subequations}
From these equations, we obtain
\begin{equation} \label{modeA}
A_n''(z, t) + \frac{\omega_n^2(t)}{c^2} \varepsilon_n(z, t) A_n(z, t) = 0,
\end{equation}
where 
\begin{equation} \label{eps}
\varepsilon_n(z, t) \equiv 1 - \frac{e^2 \rho(z, t)}{\varepsilon_0 m}
\frac{1}{\omega_n^2(t) - \Omega^2}
\end{equation}
can be regarded as the dielectric constant.
Thus, in our effective Hamiltonian, the dispersion relation is included.
If the reservoir is taken into account, the Kramers-Kronig relations will
be satisfied as shown in Ref.~\cite{Huttner} for static dielectrics.

The properties of polaritons in the matter significantly depend on the
sign of the dielectric constant $\varepsilon_n$.
When $\varepsilon_n$ is negative, the wave function of the polariton
decays in the matter, and then polaritons localize between mirrors and the
energy spectrum is discrete.
This condition is given by $\Omega^2 < \omega_n^2 < \Omega^2 +
\omega_p^2$, where $\omega_p \equiv (e^2 \rho / \varepsilon_0 m)^{1/2}$ is 
the plasma frequency.
When $\varepsilon_n$ is positive, the wave function extends indefinitely
inside the matter and the energy spectrum is continuous.

Let us consider the case in which the matter is uniform, i.e., $\rho(z, t)
= \rho \theta(-z) + \rho \theta(z - L(t))$, where $\rho$ is the
polarization density in the matter and $\theta(z)$ is the Heaviside
function.
In this case, Eq.~(\ref{modeA}) can be solved, and when $\varepsilon_n$ is
negative in the matter, the solutions are given by
\begin{subequations}
\label{sol}
\begin{eqnarray} \label{sol1}
& & A_n(z, t) = \nonumber \\
& & \left\{ \begin{array}{ll}
\alpha_n(t) e^{\kappa_n(t) z} \cos \left[ \frac{k_n(t) L(t)}{2} \right] &
(z \leq 0) \\
\alpha_n(t) \cos k_n(t) \left[ z - \frac{L(t)}{2} \right] &
(0 < z < L(t)) \\
\alpha_n(t) e^{-\kappa_n(t) [z - L(t)]} \cos \left[ \frac{k_n(t) L(t)}{2}
\right] & (z \geq L(t))
\end{array} \right. \nonumber \\
& &
\end{eqnarray}
for the even number of nodes, and
\begin{eqnarray} \label{sol2}
& & A_n(z, t) = \nonumber \\
& & \left\{ \begin{array}{ll}
-\alpha_n(t) e^{\kappa_n(t) z} \sin \left[ \frac{k_n(t) L(t)}{2} \right] &
(z \leq 0) \\
\alpha_n(t) \sin k_n(t) \left[ z - \frac{L(t)}{2} \right] &
(0 < z < L(t)) \\
\alpha_n(t) e^{-\kappa_n(t) [z - L(t)]} \sin \left[ \frac{k_n(t) L(t)}{2}
\right] & (z \geq L(t))
\end{array} \right. \nonumber \\
& &
\end{eqnarray}
\end{subequations}
for the odd number of nodes, where $k_n \equiv \omega_n / c$, $\kappa_n
\equiv |\varepsilon_n|^{1/2} k_n$, and we take the label $n$ to be the
number of nodes.
(The solutions for the continuous spectrum are given in
Appendix~\ref{app:1D}.)
From the orthonormal relation $\int dz \left[ A_n(z) \Pi_{n'}(z) - Y_n(z)
X_{n'}(z) \right] = -\delta_{nn'}$, the normalization constant $\alpha_n$
is obtained by
\begin{equation}
\alpha_n^2(t) = \left[ \varepsilon_0 \left( \frac{L(t)}{2} +
\frac{1}{\kappa_n(t)} \frac{\omega_n^2(t)}{\omega_n^2(t) - \Omega^2}
\right) \right]^{-1},
\end{equation}
where the sign of $\alpha_n$ is taken to be $A_n(0, t) > 0$.
The eigenvalues $\omega_n(t)$ are determined so that the solutions are
smoothly connected at $z = 0$ and $z = L(t)$, giving the eigenvalue
equation
\begin{equation} \label{evaleq}
\tan \frac{\omega_n(t) L(t)}{2 c} = \pm |\varepsilon_n(t)|^{\pm 1/2} = \pm
\left| 1 - \frac{\omega_p^2}{\omega_n^2(t) - \Omega^2} \right|^{\pm 1 /
2},
\end{equation}
where the signs $+$ and $-$ correspond to the solutions (\ref{sol1}) and
(\ref{sol2}), respectively.
Using Eqs.~(\ref{F1}), (\ref{F}), (\ref{sol}), (\ref{evaleq}), and
$\partial \rho(z, t) / \partial t = -\rho \dot L(t) \delta(z - L(t))$, we
obtain the coefficient (\ref{coeff}) as
\begin{subequations} \label{coeff1d}
\begin{equation}
C_{nn}(t) = -\frac{1}{8} \varepsilon_0 \dot L(t) \alpha_n^2(t),
\end{equation}
and
\begin{eqnarray}
C_{nn'} & = & \frac{1}{2} \varepsilon_0 \dot L \frac{(-1)^{n + n'}
\alpha_n \alpha_{n'}}{\omega_n^2 - \omega_{n'}^2}
\sqrt{\frac{\omega_{n'}}{\omega_n}} \frac{1}{\kappa_n + \kappa_{n'}}
\nonumber \\
& & \times \left( \kappa_{n'} \omega_n^2 \sqrt{\frac{1 +
|\varepsilon_n|}{1 + |\varepsilon_{n'}|}} + \kappa_n \omega_{n'}^2
\sqrt{\frac{1 + |\varepsilon_{n'}|}{1 + |\varepsilon_n|}} \right)
\nonumber \\
& &
\end{eqnarray}
\end{subequations}
for $n \neq n'$, where we omit the argument $t$ for brevity.

First we consider the case in which the time scale of mirror motion is
much larger than the inverse of the plasma frequency $\omega_p^{-1}$.
In this case, the transition between the discrete and continuous spectrum
can be ignored, and then the continuous spectrum is irrelevant.
In order to see the relation between our result and Eq.~(\ref{HEFFLAW}),
we consider the case of metal, which is obtained by setting $\Omega =
0$.
The coefficients (\ref{coeff1d}) reduce to
\begin{subequations} \label{coeffm}
\begin{equation}
C_{nn}(t) = -\frac{1}{4} \frac{\dot L(t)}{L(t) + \frac{2}{\kappa_n(t)}},
\end{equation}
and
\begin{eqnarray}
C_{nn'}(t) & = & \frac{(-1)^{n + n'} \omega_n(t)
\omega_{n'}(t)}{\omega_n^2(t) - \omega_{n'}^2(t)}
\sqrt{\frac{\omega_{n'}(t)}{\omega_n(t)}} \nonumber \\
& & \times \frac{\dot L(t)}{\sqrt{ \left[ L(t) + \frac{2}{\kappa_n(t)}
\right] \left[ L(t) + \frac{2}{\kappa_{n'}(t)} \right]}}
\end{eqnarray}
\end{subequations}
for $n \neq n'$.
When $\omega_n \ll \omega_p$, which corresponds to the case in which
the penetration depth of the EM field is much smaller than its wave
length, $\omega_n$ and the coefficients (\ref{coeffm}) can be expanded
with respect to $\eta(t) \equiv c / [L(t)\omega_p] \ll 1$, giving
\begin{subequations}
\begin{eqnarray}
\omega_n(t) & = & \frac{c}{L(t)} n\pi \biggl[ 1 - 2 \eta(t) + 4 \eta^2(t)
\nonumber \\
& & - 8 \left( 1 + \frac{n^2 \pi^2}{24} \right) \eta^3(t) \biggr] +
O(\eta^4), \\
C_{nn}(t) & = & -\frac{\dot L(t)}{4 L(t)} \bigl[ 1 - 2 \eta(t) + 4
\eta^2(t) \nonumber \\
& & - (8 + n^2 \pi^2) \eta^3(t) \bigr] + O(\eta^4), \\
C_{nn'}(t) & = & \frac{\dot L(t)}{L(t)} \frac{(-1)^{n + n'} n n'}{n^2 -
n^{\prime 2}} \sqrt{\frac{n'}{n}} \biggl[ 1 - 2 \eta(t) + 4 \eta^2(t)
\nonumber \\
& & - \frac{1}{3} (24 + m^2 \pi^2) \eta^3(t) \biggr] + O(\eta^4).
\end{eqnarray}
\end{subequations}
If we identify the photon operators $\hat a_n$ in $\hat H_a^{\rm eff}(t)$
[Eq.~(\ref{HEFFLAW})] as the polariton operators $\hat b_n$, we find
\begin{equation} \label{1dHeff}
\hat H_b^{\rm eff}(t) = [1 - 2\eta(t) + 4 \eta^2(t)] \hat H_a^{\rm eff}(t)
+ O(\eta^3).
\end{equation}
When we neglect the terms of order $O(\eta)$, the effective Hamiltonian
for polaritons $\hat H_b^{\rm eff}(t)$ reduces to that based on the
external boundary condition $\hat H_a^{\rm eff}(t)$, and therefore, our
method reproduces the existing results of the moving-mirror problem in the
limit of $\eta \rightarrow 0$.
It is interesting to note that $\hat H_b^{\rm eff}(t)$ is proportional to
$\hat H_a^{\rm eff}(t)$ up to the second order of $\eta$.
This physically indicates that the time scale is delayed by the factor
$\simeq 1 - 2\eta$ due to the coupling of the photon field with the matter 
field, i.e., the EM field drags electrons in the mirrors when it is
excited.
In other words, photons in the cavity are dressed by plasmons in the
cavity mirrors, forming the cavity polaritons.

\begin{figure}[tb]
\includegraphics[width=8.6cm]{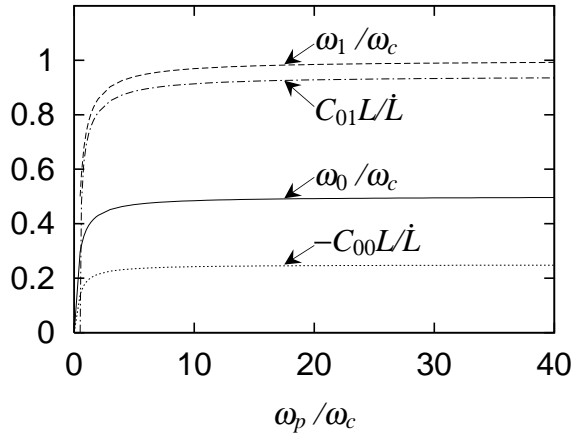}
\caption{
The eigenfrequencies of polaritons $\omega_0 / \omega_c$ (solid line) and
$\omega_1 / \omega_c$ (dashed line) and the coefficients of the effective
Hamiltonian $-C_{00} L / \dot L$ (dotted line) and $C_{01} L / \dot L$
(dot-dashed line) as functions of $\omega_p / \omega_c$, where $\omega_c
\equiv 2 \pi c / L$.
}
\label{f:omega}
\end{figure}

Figure~\ref{f:omega} shows the lowest two eigenfrequencies determined
from Eq.~(\ref{evaleq}) with $\Omega = 0$ and the coefficients
(\ref{coeffm}) for $n = 0$ and $n' = 1$ as functions of the plasma
frequency normalized by $\omega_c \equiv 2 \pi c / L$.
We find that the eigenfrequencies and the coefficients become small when
$\omega_p$ is comparable to $\omega_c$, and they go to the asymptotic
values for $\omega_p \gg \omega_c$.
For example, in the case of a superconducting microwave cavity, the plasma
frequency is in the X-ray region and $\omega_c$ is in the microwave
region, and therefore the correction to the ideal metal can be neglected.
However, this correction might be significant if $\omega_p$ becomes small
due to, e.g., decrease of the carrier density.

When the time scale of mirror motion is comparable to $\omega_p^{-1}$,
transition between discrete and continuous spectrum occurs.
The transition is significant when the mirror vibrates at the frequency
$\omega_M > \omega_p - \omega_n$, which results in decay of the polaritons
in the $n$th mode into continuous spectrum, namely, photons leak out of
the cavity.
When the mirror moves as $L(t) = L_0 + \ell \sin \omega_M t$ with $\ell \ll
L_0$, the time-dependent part of the effective Hamiltonian can be written
by $\hat V e^{i \omega_M t} + \hat V^\dagger e^{-i \omega_M t}$, where
$\hat V$ is obtained by replacing $\dot L(t)$ with $\ell \omega_M / 2$ in
$\hat H_b^{\rm eff}$.
Using Fermi's golden rule, the decay rate of a photon in the $n$th mode 
is estimated to be
\begin{equation} \label{R}
R_n(\omega_M) = \frac{2}{c\hbar^2} \sum_{i=1,2} |V_{kn}^{(i)}|^2,
\end{equation}
where $V_{kn}^{(i)}$ is the matrix element of $\hat V$ with respect to the 
$n$th mode in the discrete spectrum (\ref{sol}) and the mode labeled by
$k = (\omega_n + \omega_M) / c$ and $i = 1,2$ in the continuous spectrum
(\ref{Ak}).
Here we took the wave number $k$ between mirrors as the mode index in the
continuous spectrum.
The explicit form of $V_{kn}^{(i)}$ is given in Appendix~\ref{app:1D}.
We can show that when $\Omega = 0$ and $\omega_n \ll \omega_p \ll
\omega_M$ the decay rate (\ref{R}) reduces to $R_n(\omega_M) \simeq n
\ell^2 \omega_p^2 / (4 L_0^2 \omega_M)$.

In the effective Hamiltonian $\hat H_a^{\rm eff}(t)$, all the energy
levels are commensurate ($\omega_n = n \pi / L_0$), and the created
photons make transition to higher levels unlimitedly as $\omega_n
\rightarrow \omega_{2n} \rightarrow \cdots$ due to the resonance.
In our one-dimensional effective Hamiltonian $\hat H_b^{\rm eff}(t)$
[Eq.~(\ref{1dHeff})], on the other hand,
the transition stops at $\sim \omega_p$ due to the incommensurate energy
levels, or the decay into the continuous spectrum occurs, and thus the
resonant enhancement of the DCE is to be naturally suppressed at $\sim
\omega_p$.

\section{Conclusions}
\label{s:conclusion}

We formulated the DCE in terms of microscopic field-matter theory, in
which the EM field and the polarization field in the matter are treated on
an equal footing.
This enabled us to study the DCE without boundary conditions and without
changing the Hilbert space.
We derived the effective Hamiltonian for polaritons with moving matter,
and applied it to the one-dimensional cavity with a moving mirror.
We obtained the corrections to the results based on the external boundary
conditions: the time scale of the dynamics is delayed when the plasma
frequency is comparable to the resonant frequency of the cavity.
This effect is attributed to the fact that the photons in the cavity are
dressed by the electrons in the mirrors.

Finally, we comment on possibility of experimental observation of the DCE.
This effect has not been demonstrated in laboratories yet, since time
scale of the phenomena is extremely fast.
An efficient way to observe the DCE is to accumulate photons in a cavity
by vibrating the mirror surface of the cavity at twice the resonant
frequency~\cite{Dodonov96}.
However, the resonant frequency of the high-Q cavity is typically $\gtrsim 
10$ GHz, and it is quite difficult to excite oscillation at such a high
frequency.
One possibility to overcome this obstacle might be to slow down the speed
of light.
Using the electromagnetically induced transparency in an ultracold atomic
gas, the speed of light can be reduced to $\sim 10 \; {\rm m} / {\rm s}$
in the regime of visible light~\cite{Hau}.
If this technique can be applied to much lower frequency, the resonant
frequency of the cavity is significantly reduced, which enables us to
resonantly vibrate the cavity wall to observe the DCE.

\appendix

\section{Derivation of the effective Hamiltonian (\protect\ref{HEFFLAW})}
\label{app:Law}

In order to make this paper self-contained, we derive the effective
Hamiltonian (\ref{HEFFLAW}) for the DCE based on the external boundary
condition following the method in Ref.~\cite{Law}.
We restrict ourselves to the moving-mirror problem, while in
Ref.~\cite{Law} change of the dielectric constant is also considered.

The equations of motion are given by
\begin{subequations} \label{app:eom}
\begin{eqnarray}
\frac{\partial}{\partial t} \hat A(z, t) & = & -\hat E(z, t), \\
\frac{\partial}{\partial t} \hat E(z, t) & = & -c^2
\frac{\partial^2}{\partial z^2} \hat A(z, t),
\end{eqnarray}
\end{subequations}
and the boundary conditions are $\hat A(0, t) = \hat A(L(t), t) = 0$.
The field operators are expanded as
\begin{widetext}
\begin{subequations} \label{app:expand}
\begin{eqnarray}
\hat A(z, t) & = & \sum_n \sqrt{\frac{\hbar}{2 \varepsilon_0 \omega_n(t)}}
\phi_n(z, t) [\hat a_n(t) + \hat a_n^\dagger(t)], \\
\hat E(z, t) & = & \sum_n i \sqrt{\frac{\hbar \omega_n(t)}{2 \varepsilon_0}}
\phi_n(z, t) [\hat a_n(t) - \hat a_n^\dagger(t)],
\end{eqnarray}
\end{subequations}
where $\omega_n(t) \equiv n \pi c / L(t)$ and
\begin{equation} \label{app:phi}
\phi_n(z, t) \equiv \sqrt{\frac{2}{L(t)}} \sin \left( \frac{n \pi z}{L(t)}
\right).
\end{equation}
From Eqs.~(\ref{app:expand}) and (\ref{app:phi}), the annihilation
operator is written by
\begin{equation} \label{app:a}
\hat a_n(t) = \int_0^{L(t)} dz \frac{1}{2} \phi_n(z, t) \left
[ \sqrt{\frac{2 \varepsilon_0 \omega_n(t)}{\hbar}} \hat A(z, t) - i
\sqrt{\frac{2 \varepsilon_0}{\hbar \omega_n(t)}} \hat E(z, t) \right].
\end{equation}
Using Eqs.~(\ref{app:eom}), time derivative of Eq.~(\ref{app:a}) reads
\begin{eqnarray} \label{app:dadt}
& & \frac{d \hat a_n(t)}{dt} \nonumber \\
& = & \int_0^{L(t)} dz \frac{1}{2}
\phi_n(z, t) \left[ -\sqrt{\frac{2 \varepsilon_0 \omega_n(t)}{\hbar}} \hat
E(z, t) + i \sqrt{\frac{2 \varepsilon_0}{\hbar \omega_n(t)}} c^2
\frac{\partial^2}{\partial z^2} \hat A(z, t) \right]
- \frac{\dot L(t)}{L(t)} \int_0^{L(t)} dz \frac{1}{2} \phi_n(z, t)
\sqrt{\frac{2 \varepsilon_0 \omega_n(t)}{\hbar}} \hat A(z, t) \nonumber \\
& & - \frac{\dot L(t) n \pi}{L^2(t)} \int_0^{L(t)} dz \frac{1}{2}
\sqrt{\frac{2}{L(t)}} z \cos \left( \frac{n \pi z}{L(t)} \right) \left[
\sqrt{\frac{2 \varepsilon_0 \omega_n(t)}{\hbar}} \hat A(z, t) - i
\sqrt{\frac{2 \varepsilon_0}{\hbar \omega_n(t)}} \hat E(z, t) \right]
\nonumber \\
& = & -i \omega_n(t) \hat a_n(t) - \frac{\dot L(t)}{2 L(t)} \hat
a_n^\dagger(t)
-\frac{\dot L(t)}{L(t)} \sum_{n \neq n'} \frac{(-1)^{n + n'} nn'}{n^2
- n^{\prime 2}} \left[ \left( \sqrt{\frac{n}{n'}} + \sqrt{\frac{n'}{n}}
\right) \hat a_{n'}(t) + \left( \sqrt{\frac{n}{n'}} - \sqrt{\frac{n'}{n}}
\right) \hat a_{n'}^\dagger(t) \right].
\end{eqnarray}
\end{widetext}
Thus we find that the effective Hamiltonian (\ref{HEFFLAW}) gives this
time evolution (\ref{app:dadt}) by the Heisenberg equation.

\section{Orthogonality of the mode functions}
\label{app:orth}

We give a proof of the orthogonality of the mode functions in
Eq.~(\ref{orthonormal}).
We consider the integral
\begin{equation} \label{appint}
I_{nn'} \equiv \int d{\bf r} \left[ {\bf A}_{n'}({\bf r}) \cdot
{\bf\Pi}_n({\bf r}) - {\bf Y}_{n'}({\bf r}) \cdot {\bf X}_n({\bf r})
\right].
\end{equation}
Substituting Eqs.~(\ref{mode}b) and (\ref{mode}c) into ${\bf\Pi}_n$ and
${\bf X}_n$ in $\omega_n^2 I_{nn'}$, we find
\begin{equation} \label{appeq1}
\omega_n^2 I_{nn'} = \omega_{n'}^2 I_{n'n}.
\end{equation}
On the other hand, substituting Eqs.~(\ref{mode}a) and (\ref{mode}d) into
${\bf A}_{n'}$ and ${\bf Y}_{n'}$ in $I_{nn'}$, and using $\nabla \cdot
{\bf\Pi}_n = 0$, we find 
\begin{equation} \label{appeq2}
I_{nn'} = I_{n'n}.
\end{equation}
From Eqs.~(\ref{appeq1}) and (\ref{appeq2}), we obtain $(\omega_n^2 -
\omega_{n'}^2) I_{nn'} = 0$, i.e., $I_{nn'} = 0$ for $\omega_n^2 \neq
\omega_{n'}^2$.

\section{Calculations for continuous spectrum in one-dimension}
\label{app:1D}

When $\varepsilon_n$ is positive in the matter, the energy spectrum is
continuous, and we use the wave number $k$ between mirrors as the mode
index.
There are two independent solutions of Eq.~(\ref{modeA}) for given $k$ as
\begin{widetext}
\begin{subequations}
\label{Ak}
\begin{eqnarray}
A_k^{(1)}(z) & = & \left\{ \begin{array}{ll}
\alpha_k \left( \frac{k}{\kappa} \cos\frac{kL}{2} \sin\kappa z -
\sin\frac{kL}{2} \cos\kappa z \right) & (z \leq 0) \\
\alpha_k \sin k(z - L / 2) & (0 < z < L) \\
\alpha_k \left[ \frac{k}{\kappa} \cos\frac{kL}{2} \sin\kappa (z - L) +
\sin\frac{kL}{2} \cos\kappa (z - L) \right] & (z \geq L),
\end{array} \right. \\
A_k^{(2)}(z) & = & \left\{ \begin{array}{ll}
\beta_k \left( \frac{k}{\kappa} \sin\frac{kL}{2} \sin\kappa z +
\cos\frac{kL}{2} \cos\kappa z \right) & (z \leq 0) \\
\beta_k \cos k(z - L / 2) & (0 < z < L) \\
\beta_k \left[ -\frac{k}{\kappa} \sin\frac{kL}{2} \sin\kappa (z - L) +
\cos\frac{kL}{2} \cos\kappa (z - L) \right] & (z \geq L),
\end{array} \right.
\end{eqnarray}
\end{subequations}
\end{widetext}
where $\kappa = \varepsilon_k^{1/2} k$, and the normalization constants
\begin{subequations}
\begin{eqnarray}
\alpha_k & = & \left[ \pi \varepsilon_0 \sqrt{\varepsilon_k} \left(
\frac{1}{\varepsilon_k} \cos^2 \frac{kL}{2} + \sin^2 \frac{kL}{2} \right)
\right]^{-1/2}, \nonumber \\
& & \\
\beta_k & = & \left[ \pi \varepsilon_0 \sqrt{\varepsilon_k} \left(
\frac{1}{\varepsilon_k} \sin^2 \frac{kL}{2} + \cos^2 \frac{kL}{2} \right)
\right]^{-1/2}, \nonumber \\
& &
\end{eqnarray}
\end{subequations}
are determined from the orthonormal relation
\begin{equation}
\int dz [A_k^{(i)}(z) \Pi_{k'}^{(j)}(z) - Y_k^{(i)}(z)
X_{k'}^{(j)}(z)] = -\delta(k - k') \delta_{ij}.
\end{equation}
We can show that the functions (\ref{sol}) and (\ref{Ak}) are orthogonal
each other.
The matrix element $V_{kn}^{(i)}$ in Eq.~(\ref{R}) is obtained by the
straightforward calculation as
\begin{eqnarray}
V_{kn}^{(1)} & = & (-1)^n \frac{i \hbar \varepsilon_0 \ell \omega_M}{4}
\sqrt{\frac{k}{k_n}} \nonumber \\
& & \times \frac{\alpha_k |\alpha_n| k_p (k + k_n)}{(\kappa^2 +
\kappa_n^2)(k^2 - k_0^2)(k_n^2 - k_0^2)^{1/2}} \nonumber \\
& & \times \left( k_0^2 \kappa_n \cos
\frac{kL}{2} - k k_n^2 \sin \frac{kL}{2} \right),
\end{eqnarray}
where $k_0 \equiv \Omega / c$, $k_p \equiv \omega_p / c$, and
$V_{kn}^{(2)}$ is obtained by the replacement $\alpha_k \rightarrow
\beta_k$, $\cos kL / 2 \rightarrow -\sin kL / 2$, and $\sin kL / 2
\rightarrow \cos kL / 2$.


\begin{thebibliography}{}

\bibitem{Casimir}
H. B. G. Casimir, Proc. K. Ned. Akad. Wet. {\bf 51}, 793 (1948).
For review see, for example, G. Plunien, B. M\"{u}ller, and W. Greiner,
Phys. Rep. {\bf 134}, 87 (1986).

\bibitem{Parker}
L. Parker, Phys. Rev. Lett. {\bf 21}, 562 (1968); Phys. Rev. {\bf 183},
1057 (1969).

\bibitem{Moore}
G. T. Moore, J. Math. Phys. {\bf 11}, 2679 (1970).

\bibitem{Fulling}
S. A. Fulling and P. C. W. Davies, Proc. R. Soc. London A {\bf 348}, 393
(1976).

\bibitem{Castagnino}
See, for example, M. Castagnino and R. Ferraro, Ann. Phys. (N.Y.) {\bf
154}, 1 (1984).

\bibitem{Dodonov89}
V. V. Dodonov, A. B. Klimov, and V. I. Man'ko, Phys. Lett. A {\bf 142},
511 (1989).

\bibitem{Jaekel}
M. T. Jaekel and S. Reynaud, J. Phys. I France {\bf 2}, 149 (1992).

\bibitem{Neto}
P. A. Maia Neto and S. Reynaud, Phys. Rev. A {\bf 47}, 1639 (1993);
P. A. Maia Neto, J. Phys. A {\bf 27}, 2167 (1994).

\bibitem{Dodonov90}
V. V. Dodonov, A. B. Klimov, and V. I. Man'ko, Phys. Lett. A {\bf 149},
225 (1990).

\bibitem{Sarkar}
S. Sarkar, Quantum Opt. {\bf 4}, 277 (1992).

\bibitem{Razavy}
M. Razavy and J. Terning, Phys. Rev. D {\bf 31}, 307 (1985).

\bibitem{Law}
C. K. Law, Phys. Rev. A {\bf 49}, 433 (1994).

\bibitem{Law95}
C. K. Law, Phys. Rev. A {\bf 51}, 2537 (1995).

\bibitem{Schutz}
R. Sch\"utzhold, G. Plunien, and G. Soff, Phys. Rev. A {\bf 57}, 2311
(1998).

\bibitem{Yab}
E. Yablonovitch, Phys. Rev. Lett. {\bf 62}, 1742 (1989).

\bibitem{Dodonov93}
V. V. Dodonov, A. B. Klimov, and D. E. Nikonov, Phys. Rev. A {\bf 47},
4422 (1993).

\bibitem{Okushima}
T. Okushima and A. Shimizu, Jpn. J. Appl. Phys. {\bf 34}, 4508 (1995);
A. Shimizu, T. Okushima, and K. Koshino, Materials Science and Engineering 
B{\bf 48}, 66 (1997).

\bibitem{Saito}
H. Saito and H. Hyuga, J. Phys. Soc. Jpn. {\bf 65}, 1139 (1996); {\em
ibid.} {\bf 65} 3513 (1996).

\bibitem{Barton}
G. Barton and C. Eberlein, Ann. Phys. (N.Y.) {\bf 227}, 222 (1993).

\bibitem{Salamone}
G. M. Salamone and G. Barton, Phys. Rev. A {\bf 51}, 3506 (1995).

\bibitem{Barton95}
G. Barton and A. Calogeracos, Ann. Phys. (N.Y.) {\bf 238}, 227 (1995);

\bibitem{Barton96}
G. Barton and C. A. North, Ann. Phys. (N.Y.) {\bf 252}, 72 (1996).

\bibitem{Gutig}
R. G\"utig and C. Eberlein, J. Phys. A {\bf 31}, 6819 (1998).

\bibitem{Eberlein99}
C. Eberlein, J. Phys. A {\bf 32}, 2583 (1999).

\bibitem{Dodonov98}
V. V. Dodonov, A. B. Klimov, and D. E. Nikonov, J. Math. Phys. {\bf 34},
2742 (1993).

\bibitem{Lambrecht}
A. Lambrecht, M. T. Jaekel, and S. Reynaud, Europhys. Lett. {\bf 43}, 147
(1998).

\bibitem{Plunien}
G. Plunien, R. Sch\"{u}tzhold, and G. Soff, Phys. Rev. Lett. {\bf 84},
1882 (2000).

\bibitem{Schwinger}
J. Schwinger, Proc. Natl. Acad. Sci. USA {\bf 89}, 4091 (1992); {\bf 90}, 
958 (1993); {\bf 90}, 2105 (1993); {\bf 90}, 4505 (1993); {\bf 90}, 7285
(1993); {\bf 91}, 6473 (1994).

\bibitem{Eberlein}
C. Eberlein, Phys. Rev. Lett. {\bf 76}, 3842 (1996); Phys. Rev. A {\bf
53}, 2772 (1996).

\bibitem{Dodonov96}
V. V. Dodonov and A. B. Klimov, Phys. Rev. A {\bf 53}, 2664 (1996).

\bibitem{Dodonov01}
For recent review, V. V. Dodonov, quant-ph/0106081.

\bibitem{Koashi}
M. Koashi and M. Ueda, Phys. Rev. A{\bf 58}, 2699 (1998).

\bibitem{Fano}
U. Fano, Phys. Rev. {\bf 103}, 1202 (1956).

\bibitem{Hopfield}
J. J. Hopfield, Phys. Rev. {\bf 112}, 1555 (1958).

\bibitem{Glauber}
R. J. Glauber and M. Lewenstein, Phys. Rev. A {\bf 43}, 467 (1991).

\bibitem{Huttner}
B. Huttner, J. J. Baumberg, and S. M. Barnett, Europhys. Lett. {\bf 16},
177 (1991); B. Huttner and S. M. Barnett, Europhys. Lett. {\bf 18}, 487
(1992); Phys. Rev. A {\bf46}, 4306 (1992).

\bibitem{Matloob}
R. Matloob, R. Loudon, S. M. Barnett, and J. Jeffers, Phys. Rev. A {\bf
52}, 4823 (1995).

\bibitem{Messiah}
A. Messiah, {\em M\'{e}canique Quantique} (Dunod, Paris, 1964).

\bibitem{Landau}
L. D. Landau, E. M. Lifshitz, and L. P. Pitaevskii, {\it Electrodynamics
of Continuous Media} (Pergamon, Oxford, 1984).

\bibitem{Hau}
L. V. Hau, S. E. Harris, Z. Dutton, and C. H. Behroozi, Nature {\bf 397},
594 (1999).

\end{thebibliography}
\end{document}